\documentclass[preprint]{aastex631} 

\usepackage{textcomp}  
\usepackage{comment}
\usepackage{subfigure}
\usepackage{physics}
\usepackage{color}

\received{August 18, 2021}
\revised{October 15, 2021}
\accepted{October, 2021}
\submitjournal{ApJ}

\shorttitle{HighEneAntivKam2021}
\shortauthors{KamLAND collaboration}

\graphicspath{{./}{figures/}}

\begin{document}
\title{Limits on astrophysical antineutrinos with the KamLAND experiment}

\correspondingauthor{S.~Obara}
\email{shuhei.obara.d4@tohoku.ac.jp}

\newcommand{\tohoku}{\affiliation{Research Center for Neutrino Science, Tohoku University, Sendai 980-8578, Japan}}
\newcommand{\fris}{\affiliation{Frontier Research Institute for Interdisciplinary Sciences, Tohoku University, Sendai 980-8578, Japan}}
\newcommand{\gppu}{\affiliation{Graduate Program on Physics for the Universe, Tohoku University, Sendai 980-8578, Japan}}
\newcommand{\tohokuRigaku}{\affiliation{Department of Physics, Tohoku University, Sendai 980-8578, Japan}}
\newcommand{\ipmu}{\affiliation{Institute for the Physics and Mathematics  of the Universe, The University of Tokyo, Kashiwa 277-8568, Japan}}
\newcommand{\osaka}{\affiliation{Graduate School of Science, Osaka University, Toyonaka, Osaka 560-0043, Japan}}   
\newcommand{\osakarcnp}{\affiliation{Research Center for Nuclear Physics (RCNP), Osaka University, Ibaraki, Osaka 567-0047, Japan}}
\newcommand{\tokushima}{\affiliation{Graduate School of Advanced Technology and Science, Tokushima University, Tokushima 770-8506, Japan}}
\newcommand{\tokushimaGakusei}{\affiliation{Graduate School of Integrated Arts and Sciences, Tokushima University, Tokushima 770-8502, Japan}}
\newcommand{\kyoto}{\affiliation{Department of Physics, Kyoto University, Kyoto 606-8502, Japan}}
\newcommand{\lbl}{\affiliation{Nuclear Science Division, Lawrence Berkeley National Laboratory, Berkeley, CA 94720, USA}}
\newcommand{\hawaii}{\affiliation{Department of Physics and Astronomy, University of Hawaii at Manoa, Honolulu, HI 96822, USA}}
\newcommand{\mituniv}{\affiliation{Massachusetts Institute of Technology, Cambridge, MA 02139, USA}}
\newcommand{\bu}{\affiliation{Boston University, Boston, MA 02215, USA}}
\newcommand{\tennessee}{\affiliation{Department of Physics and Astronomy,  University of Tennessee, Knoxville, TN 37996, USA}}
\newcommand{\tunl}{\affiliation{Triangle Universities Nuclear Laboratory, Durham, NC 27708, USA}}    
\newcommand{\chapehill}{\affiliation{The University of North Carolina at Chapel Hill, Chapel Hill, NC 27599, USA}}

\newcommand{\northcarolina}{\affiliation{North Carolina Central University, Durham, NC 27701, USA}}
\newcommand{\duke}{\affiliation{Physics Department at Duke University, Durham, NC 27705, USA}}
\newcommand{\seattle}{\affiliation{Center for Experimental Nuclear Physics and Astrophysics, University of Washington, Seattle, WA 98195, USA}}
\newcommand{\nikhef}{\affiliation{Nikhef and the University of Amsterdam, Science Park,  Amsterdam, the Netherlands}}
\newcommand{\virginia}{\affiliation{Center for Neutrino Physics, Virginia Polytechnic Institute and State University, Blacksburg, VA 24061, USA}}

\author{S.~Abe}\tohoku
\author{S.~Asami}\tohoku
\author{A.~Gando}\tohoku
\author{Y.~Gando}\tohoku
\author{T.~Gima}\tohoku 
\author{A.~Goto} \tohoku
\author{T.~Hachiya}\tohoku
\author{K.~Hata} \tohoku
\author{S.~Hayashida} \altaffiliation{Present address: Imperial College London, Department of Physics, Blackett Laboratory, London SW7 2AZ, UK} \tohoku
\author{K.~Hosokawa} \tohoku
\author{K.~Ichimura} \tohoku  
\author{S.~Ieki} \tohoku
\author{H.~Ikeda}\tohoku
\author{K.~Inoue}\tohoku \ipmu 
\author{K.~Ishidoshiro}\tohoku
\author{Y.~Kamei} \tohoku
\author{N.~Kawada} \tohoku
\author{Y.~Kishimoto} \tohoku \ipmu
\author{T.~Kinoshita} \tohoku 
\author{M.~Koga}\tohoku \ipmu 
\author{N.~Maemura}\tohoku
\author{T.~Mitsui}\tohoku
\author{H.~Miyake}\tohoku
\author{K.~Nakamura}\tohoku 
\author{K.~Nakamura}\tohoku 
\author{R.~Nakamura}\tohoku
\author{H.~Ozaki}\tohoku \gppu
\author{T.~Sakai} \tohoku 
\author{H.~Sambonsugi}\tohoku
\author{I.~Shimizu}\tohoku
\author{Y.~Shirahata} \tohoku
\author{J.~Shirai}\tohoku
\author{K.~Shiraishi}\tohoku
\author{A.~Suzuki}\tohoku
\author{Y.~Suzuki}\tohoku 
\author{A.~Takeuchi}\tohoku
\author{K.~Tamae}\tohoku
\author{K.~Ueshima} \altaffiliation{Present address: National Institutes for Quantum and Radiological Science and Technology (QST), Hyogo 679-5148, Japan} \tohoku 
\author{Y.~Wada}\tohoku
\author{H.~Watanabe}\tohoku
\author{Y.~Yoshida} \tohoku
\author{S.~Obara}\fris 
\author{A.~K.~Ichikawa} \tohokuRigaku
\author{A.~Kozlov} \altaffiliation{Present address: National Research Nuclear University ``MEPhI'' (Moscow Engineering Physics Institute), Moscow, 115409, Russia} \ipmu 
\author{D.~Chernyak} \altaffiliation{Present address: Department of Physics and Astronomy, University of Alabama, Tuscaloosa, AL 35487, USA} \ipmu 
\author{Y.~Takemoto} \altaffiliation{Present address: Kamioka Observatory, Institute for Cosmic-Ray Research, The University of Tokyo, Hida, Gifu 506-1205, Japan} \osakarcnp 
\author{S.~Yoshida}\osakarcnp
\author{S.~Umehara}\osaka
\author{K.~Fushimi}\tokushima
\author{S.~Hirata} \tokushimaGakusei
\author{K.~Z.~Nakamura}\kyoto
\author{M.~Yoshida}\kyoto 
\author{B.~E.~Berger}\lbl \ipmu
\author{B.~K.~Fujikawa}\lbl \ipmu
\author{J.~G.~Learned}\hawaii
\author{J.~Maricic}\hawaii
\author{S.~N.~Axani}\mituniv
\author{L.~A.~Winslow}\mituniv
\author{Z.~Fu}\mituniv
\author{J.~Ouellet}\mituniv
\author{Y.~Efremenko}\tennessee \ipmu
\author{H.~J.~Karwowski}\tunl \chapehill
\author{D.~M.~Markoff}\tunl \northcarolina
\author{W.~Tornow}\tunl \duke \ipmu
\author{A.~Li}\chapehill
\author{J.~A.~Detwiler}\seattle \ipmu
\author{S.~Enomoto}\seattle \ipmu
\author{M.~P.~Decowski}\nikhef \ipmu
\author{C.~Grant}\bu
\author{T.~O'Donnell}\virginia
\author{S.~Dell'Oro}\virginia

\collaboration{99}{(KamLAND Collaboration)}

\begin{abstract}
We report on a search for electron antineutrinos ($\bar{\nu}_e$) from astrophysical sources in the neutrino energy range $8.3$ to $30.8\,{\mathrm{MeV}}$ with the KamLAND detector.
In an exposure of $6.72$\,kton-year of the liquid scintillator,
we observe $18$ candidate events via the inverse beta decay reaction.
Although there is a large background uncertainty from neutral current atmospheric neutrino interactions, we find no significant excess over background model predictions.
Assuming several supernova relic neutrino spectra, we give upper flux limits of $60$--$110\,{\mathrm{cm}^{-2}\mathrm{s}^{-1}}$ ($90\%$~CL) in the analysis range and present a model-independent flux.
We also set limits on the annihilation rates for light dark matter pairs to neutrino pairs.
These data improves on the upper probability limit of $^{8}$B solar neutrinos converting into $\bar{\nu}_e$'s, $P_{\nu_e \rightarrow \bar{\nu}_e} < 3.5\times10^{-5}$ ($90\%$~CL) assuming an undistorted $\bar{\nu}_e$ shape.
This corresponds to a solar $\bar{\nu}_e$ flux of 
$60\,{\mathrm{cm}^{-2}\mathrm{s}^{-1}}$ ($90\%$~CL)
in the analysis energy range. 
\end{abstract}


\section{Introduction} \label{sec:intro}
Underground liquid-scintillator neutrino detectors observe geo neutrinos, solar neutrinos, and reactor neutrinos below $10\,\mathrm{MeV}$ energy, in addition to the atmospheric neutrinos peak at $\order{\mathrm{GeV}}$ range.
However, other astrophysical neutrino sources also exist in our universe: from supernova explosions to hypothetical dark-matter annihilation neutrinos.
The valley of the neutrino energy spectrum between the end of reactor neutrinos and the onset of atmospheric neutrinos can be used to search for astrophysical neutrinos.
We present a search for astrophysical neutrinos in the neutrino energy between $8.3$ and $30.8\,\mathrm{MeV}$, focusing on electron antineutrinos ($\bar{\nu}_e$) from the Sun, past supernovae, and dark matter annihilation.

The Sun is the dominant source of astrophysical neutrinos, and various neutrino detectors have observed solar electron neutrinos ($\nu_e$)~\citep{PhysRevC.84.035804, PhysRevC.88.025501, PhysRevD.94.052010, PhysRevD.101.062001}.
As discussed in \citet{Malaney_1990}, antineutrinos are also produced in the Sun in comparatively small amounts from beta decays of $^{40}$K, $^{232}$Th, and $^{238}$U, and they has not been observed yet.
Solar neutrinos can be converted to antineutrinos with combined processes of the Mikheyev-Smirnov-Wolfenstein (MSW) effect~\citep{Smirnov_2005} and Resonant Spin Flavor Precession (RSFP)~\citep{PhysRevD.37.1368,AKHMEDOV198864} as discussed in \citet{AKHMEDOV20037} and \citet{PhysRevD.80.076007}.
This happens in two-step processes:
\begin{eqnarray}
    \nu_e \xrightarrow{\mathrm{MSW}} \nu_{\mu} \xrightarrow{\mathrm{RSFP}} \bar{\nu}_e, \\
    \nu_e \xrightarrow{\mathrm{RSFP}} \bar{\nu}_{\mu} \xrightarrow{\mathrm{MSW}} \bar{\nu}_e.
\end{eqnarray}
The RSFP is a neutrino helicity resonance transition similar to the MSW effect in the Sun via the neutrino magnetic moment $(\mu)$.
A simple RSFP model is excluded due to the large neutrino magnetic moment required, $\mu > 10^{-10}\,{\mu_B}$, already excluded by experiments~\citep{Beda2013, PhysRevD.96.091103}.
The $\mu_B$ is the Bohr magneton.
The combined RSFP+MSW model is still allowed.
The conversion probability is expressed as
\begin{equation}
    P(\nu_{e} \rightarrow \bar{\nu}_{e}) \simeq 1.8 \times 10^{-10} \sin^{2}{2\theta_{12}} \times \qty[\frac{\mu}{10^{-12} \mu_B} \frac{B_{T}(0.05 R_{\odot})}{10\,\mathrm{kG}}]^2,
    \label{eq:rsftprob}
\end{equation}
where $\theta_{12}$ is the neutrino mixing angle in the Pontecorvo-Maki-Nakagawa-Sakata matrix, $B_T$ is the transverse solar magnetic field in the region of neutrino production, $R_{\odot}$ is the solar radius, $\mu$ is the neutrino magnetic moment. 
Experimentally, the conversion probability for solar $^{8}$B neutrinos was studied by KamLAND~\citep{Gando_2012}, Borexino~\citep{AGOSTINI2021102509}, and Super-K~\citep{Abe:2020tyy}.

A supernova explosion is one of the largest neutrino burst events in our universe.
Supernova neutrinos from SN1987A, which occurred on 1987 February 23rd in the Large Magellanic Cloud, were detected by the water cherenkov detectors: KamiokaNDE~\citep{PhysRevLett.58.1490, PhysRevD.38.448} and IMB~\citep{PhysRevLett.58.1494, PhysRevD.37.3361}, and the Baksan scintillation detector~\citep{ALEXEYEV1988209}. 
A future nearby supernova explosion could reveal detailed information on the explosion mechanism.
At the same time, neutrinos from all the past supernovae are still traveling in our universe. 
These are called supernova relic neutrinos (SRN), and they provide the diffuse supernova neutrino flux.
The SRN energy spectrum and associated detection rates have been discussed in various models~\citep{PhysRevD.62.043001, PhysRevD.79.083013, Nakazato_2013, Nakazato_2015}.
The most stringent experimental $\bar{\nu}_e$  flux upper limit is given by Super-K~\citep{kamiokandecollaboration2021diffuse}, but no significant signal observation has been made yet.
KamLAND is able to perform a comparable search for $\bar{\nu}_e$ at around $10\,{\mathrm{MeV}}$. 
The higher neutron tagging efficiency should give an advantage over Super-K searching at the lower energy region.

Neutrinos can also be produced in the annihilation of dark matter particles.
In case of the existence of an MeV-scale light dark matter particle, its self-annihilation process might produce neutrino pairs ($\chi \chi \rightarrow \nu \bar{\nu}$) at MeV energies.
Assuming a model of MeV-scale dark matter annihilation in our galactic halo~\citep{PhysRevD.77.025025}, the $\bar{\nu}_e$ flux from dark-matter self-annihilation is given by
\begin{equation}
    \dv{\phi}{E_{\nu}} = \frac{\langle \sigma_A \vb{v} \rangle}{2}\, \mathcal{J}_{\mathrm{ave}}\, \frac{R_{\mathrm{sc}}\, \rho^2_0}{m^2_{\chi}}\, \frac{1}{3}\, \delta \qty(E_{\nu} - m_{\chi}),
    \label{eq:dmcrosssection}
\end{equation}
where $m_{\chi}$ is the dark matter mass, $\langle \sigma_A \vb{v} \rangle$ is the averaged self-annihilation cross section times the relative velocity of the annihilating particles, $\mathcal{J}_{\mathrm{ave}}$ is the angular-averaged intensity over the whole Milky Way, $R_{\mathrm{sc}} = 8.5\,\mathrm{kpc}$ is 
the distance of the Sun to the galactic center, 
and $\rho_0 = 0.3\,{\mathrm{GeV\, cm^{-3}}}$ is the local dark matter density.
Here, a factor $1/3$ is assumed for the branching ratio to the three flavors of neutrinos.
This process is also discussed in \citet{PhysRevD.98.103004} and \citet{Arguelles:2019ouk}.
In this work, we show the results for two benchmark cases of $\mathcal{J}_{\mathrm{ave}}=1.3$ and $5.0$~\citep{PhysRevD.77.025025}.

In this paper, after describing the KamLAND experiment (Section~\ref{sec:detector}), we describe the search for $\bar{\nu}_e$ with an energy range of $8.3$--$30.8\,\mathrm{MeV}$ (Section~\ref{sec:antiv}).
The main backgrounds are discussed in Section~\ref{sec:backgrounds}; these are reactor neutrinos, accidental backgrounds, spallation products, fast neutrons, and atmospheric neutrinos.
The data is interpreted in Section~\ref{sec:analysis} and 
we present the conversion probability of solar $^{8}$B antineutrinos, the SRN flux, and the dark-matter self-annihilation cross section.
We summarize those results in Section~\ref{sec:summary}.

\section{KamLAND detector} \label{sec:detector} 
\begin{figure}[tb]
    \centering
    \includegraphics[width=1.0\linewidth]{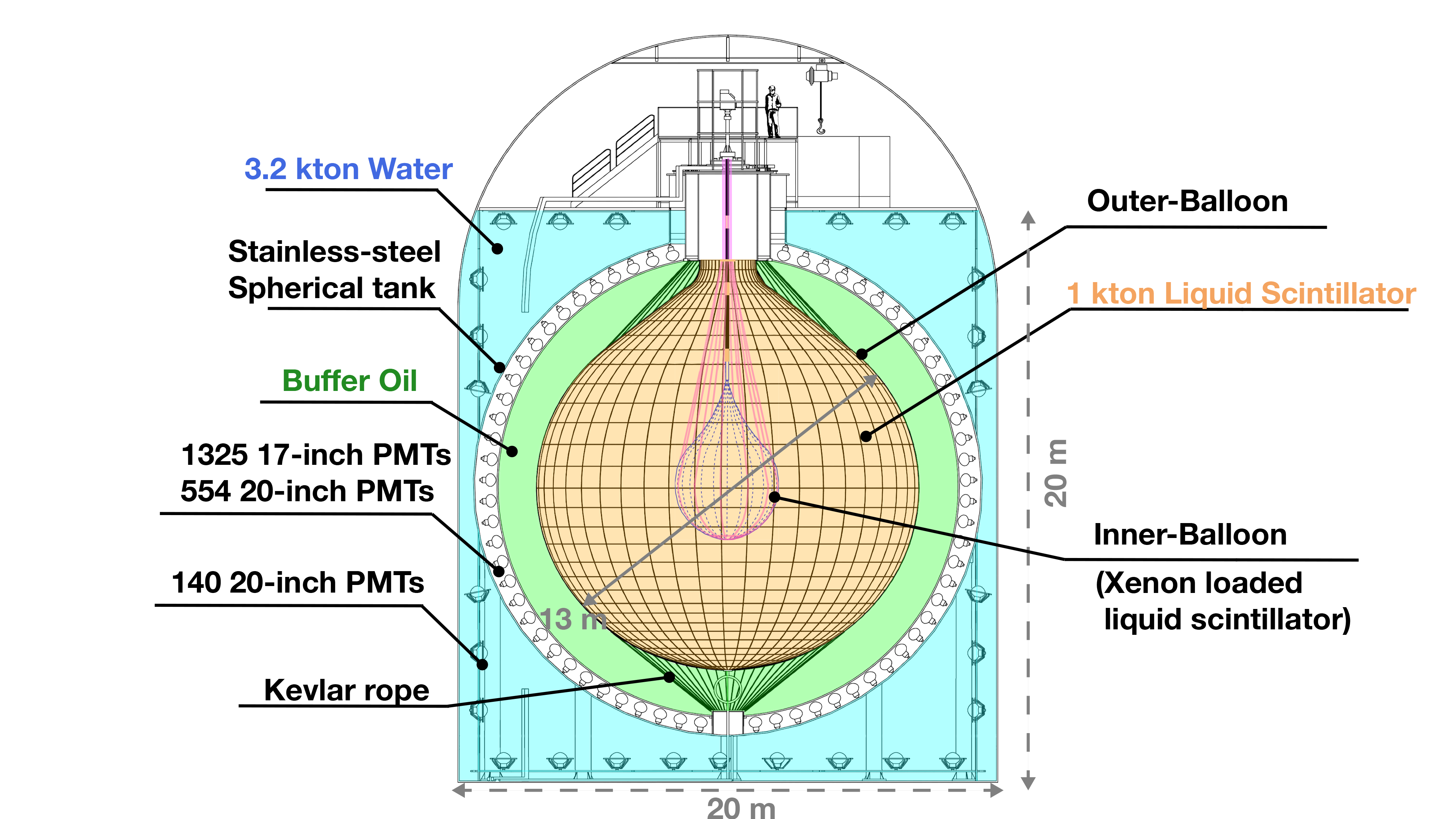}
    \caption{Schematic view of the KamLAND experiment. After the OD refurbishment campaign in 2016, the number of PMTs in the OD changed from 225 to 140~\citep{Ozaki:2016fmr}.
    The first inner balloon was installed in August 2011.}
    \label{fig:kamland}
\end{figure}

The KamLAND experiment uses ultra pure liquid scintillator to detect $\bar{\nu}_e$ via the inverse beta-decay (IBD) reaction.
The detector is located $1000\,\mathrm{m}$ underground, underneath Mt.~Ikenoyama in Kamioka Japan, corresponding to $2700\,\mathrm{m}$ water equivalent.
The cosmic muon flux is suppressed by $\sim \mathcal{O}(10^{5})$ relative to sea level.
Figure~\ref{fig:kamland} shows a schematic view of the detector.
KamLAND consists of a $18$-m diameter stainless-steel sphere tank (Inner Detector, ID) and a cylindrical vessel of water-cherenkov muon veto (outer detector, OD) surrounding the ID.
A $13$-m diameter EVOH/nylon balloon (outer-balloon) holds $1$-kton of liquid scintillator at the center of the ID.
Photon sensors, $1325$ 17-inch and $554$ 20-inch photomultiplier tubes (PMTs), sensitive to scintillation light are bolted to the inside of the stainless-steel tank.
Details of the detector design are described in~\citet{Suzuki2014}.

KamLAND data acquisition started in March 2002, and this study uses all data sets up to July 2020.
The KamLAND-Zen~400 phase of the project operated with a $154\,\mathrm{cm}$-radius nylon balloon (inner-balloon) at the center of KamLAND filled with xenon-loaded liquid scintillator from August 2011 to December 2015~\citep{PhysRevLett.117.082503}. 
The detector was further upgraded in May 2018 when KamLAND-Zen~800 started with a new $1.9\,\mathrm{m}$-radius inner-balloon~\citep{GandoY_2020, zencollaboration2021nylon}.
In this study, the inner-balloon volume is vetoed to avoid possible background contamination.
The OD system was refurbished in 2016, when the $225$ PMTs were replaced by $140$ new PMTs including $47$ higher quantum efficiency PMTs~\citep{Ozaki:2016fmr}.

The interaction vertex and energy deposition are reconstructed using the measured PMT charge and timing information. 
At low energies, the detector was calibrated using various radioactive sources: $^{60}$Co, $^{68}$Ge, $^{203}$Hg, $^{65}$Zn, $^{241}$Am$^{9}$Be, $^{137}$Cs, and $^{210}$Po$^{13}$C. 
Above $10\,\mathrm{MeV}$, the energy response is calibrated using spallation-products of $^{12}$B ($\tau = 29.1\,\mathrm{ms}$, $Q = 13.4\,\mathrm{MeV}$) and $^{12}$N ($\tau = 15.9\,\mathrm{ms}$, $Q = 17.3\,\mathrm{MeV}$).

The position-dependent energy calibration and fiducial volume determination uncertainty were studied with calibration sources positioned throughout the outer balloon volume~\citep{berger2009kamland}.
The reconstructed energy and interaction vertex resolution were determined to be $6.4\%/\sqrt{E\,({\mathrm{MeV}})}$ and $\sim12\,{\mathrm{cm}}/\sqrt{E\,({\mathrm{MeV}})}$~\citep{PhysRevD.88.033001}, respectively.
Daily stability measurements were performed using $2.2\,\mathrm{MeV}$ gamma rays emitted from spallation-induced neutron capture on protons~\citep{PhysRevC.81.025807} and spallation $^{12}$B events.
The total estimated uncertainty including the time variation of the energy scale, linearity, and uniformity is within $\pm2.0\%$ for this data set.

The primary radioactive backgrounds in the liquid scintillator are $(5.0\pm0.2)\times 10^{-18}\,\mathrm{g\,g^{-1}}$ 
of $^{238}$U 
and $(1.8\pm0.1)\times10^{-17}\mathrm{g\,g^{-1}}$ 
of $^{232}$Th~\citep{PhysRevC.92.055808}.  
These radioactive contaminants are negligibly small relative to other backgrounds in this study, such as muon spallation products and atmospheric neutrinos, and are therefore ignored.

\section{Electron antineutrino selection} \label{sec:antiv}
Electron antineutrinos are detected in KamLAND via the IBD reaction ($\bar{\nu}_e + p \rightarrow e^+ + n$) with a $1.8\,\mathrm{MeV}$ neutrino energy threshold.
IBD candidate events are selected by the delayed coincidence (DC) method: scintillation light from the positron and its annihilation gamma-rays is the prompt event, followed by a $2.2\,\mathrm{MeV}$ ($4.9\,\mathrm{MeV}$) gamma-ray from neutron capture on a proton (carbon-12) after a mean capture time of $207.5\pm2.8\,\mu\mathrm{s}$~\citep{PhysRevC.81.025807} (delayed event).
The incident neutrino energy ($E_{\nu}$) is computed from the reconstructed prompt energy ($E_{\mathrm{prompt}}$),
$E_{\nu} \simeq E_{\mathrm{prompt}} + 0.8\,{\mathrm{MeV}} + \overline{E}_n$, where $\overline{E}_n$ is the average neutron kinetic energy of $\order{10\,\mathrm{keV}}$.

The DC selection criteria between the prompt and delayed events use the prompt energy, delayed energy ($E_{\mathrm{delayed}}$), spatial distribution ($\Delta R$), and time difference ($\Delta T$); they are $7.5 < E_{\mathrm{prompt}} < 30\,\mathrm{MeV}$, $1.8 < E_{\mathrm{delayed}} < 2.6\,\mathrm{MeV}$ or $4.4 < E_{\mathrm{delayed}} < 5.6\,\mathrm{MeV}$, $\Delta R < 160\,\mathrm{cm}$, and $0.5 < \Delta T < 1000\,\mu\mathrm{s}$, respectively.
The two delayed-energy selection criteria correspond to a 2.2-MeV capture gamma-ray on a proton and a 4.9-MeV capture gamma-ray on a carbon-12. The timing difference between the prompt and delayed events are required from the 207.5-$\mu$s of neutron capture time. The spatial correlation selection is optimized from diffusion length of a thermalized neutron and delayed gamma ray. The selection efficiency is evaluated with Monte Carlo (MC) simulations described in next paragraph.
In this energy range, one of the primary backgrounds is a fast neutron (described in Sec.~\ref{sec:backgrounds}), mostly in close proximity to the ID vessel.
In order to suppress this contamination, we select a $550\,\mathrm{cm}$-radius spherical fiducial volume from the center of KamLAND, corresponding to a total number of target protons of 
$N_{p} = (4.6\pm0.1) \times 10^{31}$.

During the KamLAND-Zen~400/800 phases, the inner-balloon regions are vetoed for the delayed event in order to avoid background contamination from the xenon-loaded liquid scintillator, inner balloon body, and suspending ropes. 
The inner-balloon regions are cut from the analysis: a $250$-cm-radius spherical volume centered in the detector and a $250$-cm-radius vertical cylindrical volume in the upper-half of the detector.
For the above DC selection, the IBD detection efficiency $\epsilon_{\mathrm{IBD}}$ is estimated through MC simulations with uniformly generated neutrino events and is determined to be 
$\epsilon_{\mathrm{IBD}} = 92\% (73\%)$ for without(with) inner-balloon cut.

\begin{figure}[htbp]
    \centering
    \includegraphics[width=1.0\linewidth]{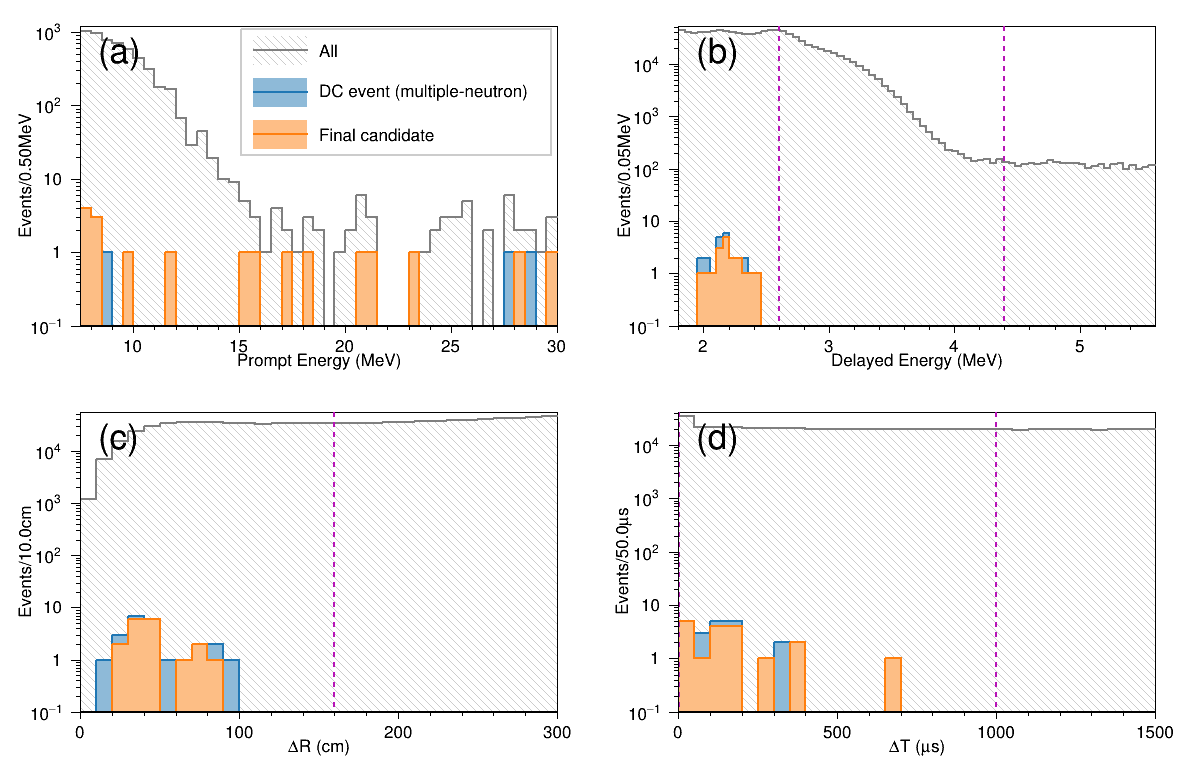}
    \caption{Event distributions for all events, after DC selection, and the final sample of $18$ events: 
    (a) prompt energy spectrum, 
    (b) delayed energy spectrum, 
    (c) spatial distribution between prompt and delayed events, 
    and (d) time difference between prompt and delayed events. 
    Vertical dashed lines correspond to cut threshold values. 
    The blue histograms include multiple-delayed neutron events and are rejected in the final candidate selection. 
    The orange histograms are the final antineutrino candidates.}
    \label{fig:CandidateSpectrum}  
\end{figure}

\begin{figure}[htbp]
    \centering
    \includegraphics[width=0.8\linewidth]{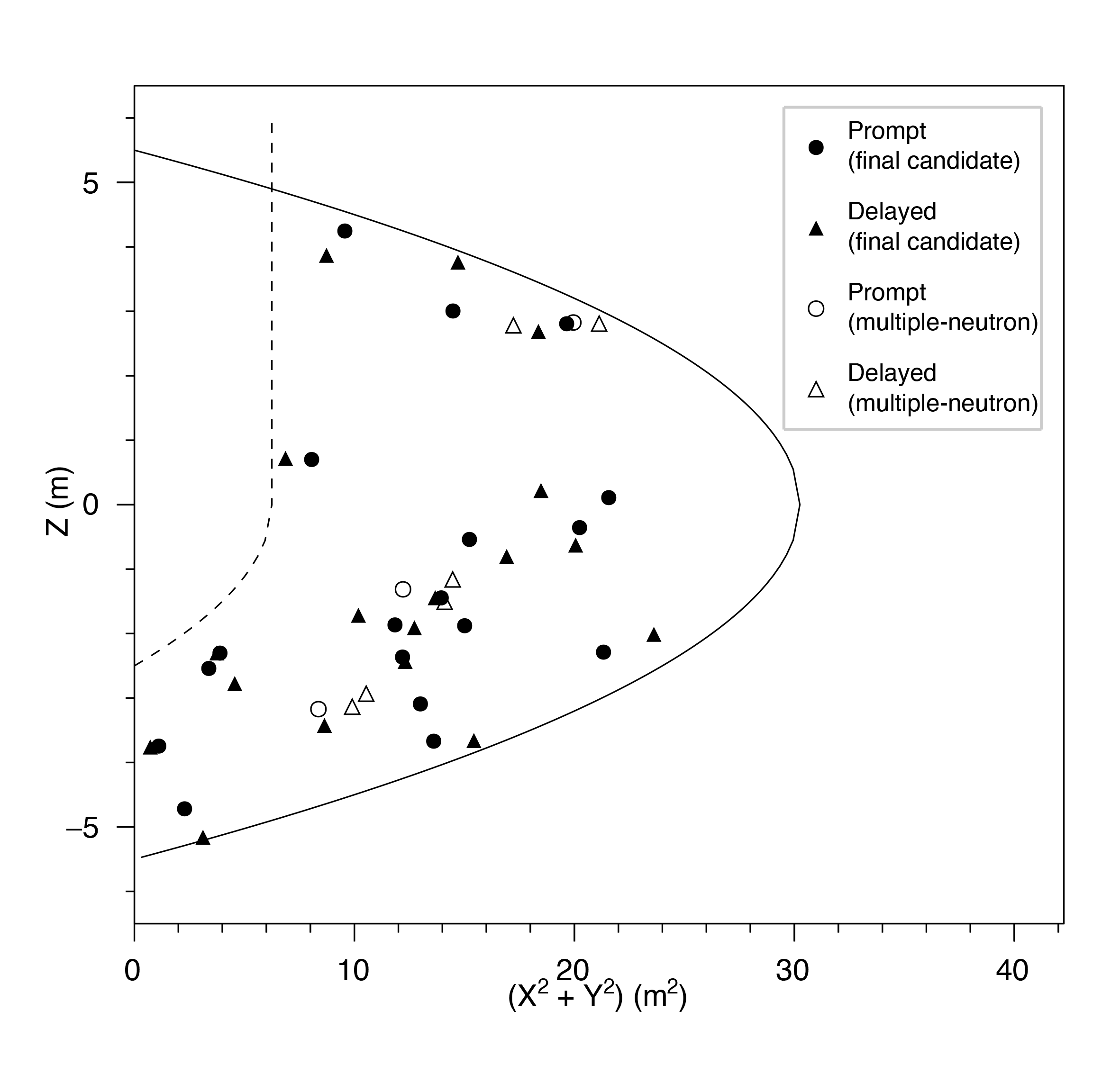}
    \caption{Position distribution after all cuts and DC selection. 
    The filled circles and triangles are the final prompt and delayed positions, respectively.  
    The unfilled circles and triangles correspond to multiple-neutron DC events. 
    The solid curve shows the fiducial radius of $550\,\mathrm{cm}$, and the dashed line represents the inner-balloon cut region.}
    \label{fig:CandidateVertex}
\end{figure}

The data period from May 2002 to July 2020 totals $4528.5$ days of livetime. 
We find $21$ DC pairs after DC selection, $3$ of them have multiple delayed events following the prompt event and are excluded from the final sample as they are likely due to the fast neutron backgrounds and/or atmospheric neutrino interactions.
Figure~\ref{fig:CandidateSpectrum} and Figure~\ref{fig:CandidateVertex} present the electron antineutrino candidate distributions.


\section{Background Estimation} \label{sec:backgrounds}
Possible backgrounds in this analysis come from reactor antineutrinos, the accidental coincidence of events, spallation products, fast neutron, and atmospheric neutrinos.
Radioactive backgrounds and reactor neutrinos were studied during the reactor- and geo-neutrino measurements at KamLAND~\citep{Gando2011, PhysRevD.88.033001}, while previous $\bar{\nu}_e$ analyses in the $\order{10-10^{3}}\,\mathrm{MeV}$ energy range~\citep{Gando_2012, asakura2015study, PhysRevD.92.052006, Abe_2021} showed that fast neutron and atmospheric neutrinos are the most challenging backgrounds above $\sim 10\,\mathrm{MeV}$.

\subsection{Reactor antineutrinos}
The location of the KamLAND detector is surrounded by $56$ Japanese nuclear power reactors. 
The reactor neutrino flux comes primarily from the beta decay of neutron-rich fragments produced in the fission of $4$ isotopes: $^{235}$U, $^{238}$U, $^{239}$Pu, and $^{241}$Pu.
For each reactor, the appropriate operational records including thermal power generation, fuel burn-up, shutdowns, and fuel reload schedule were used to calculate the fission rates.
With the reactor operation data, we have measured reactor antineutrinos between a few to several MeV~\citep{PhysRevD.83.052002, PhysRevD.88.033001}.
However, there are no measured reactor neutrino spectra in the present analysis energy range. 
Hence, we use the reactor neutrino spectra assuming polynomial functions based on the results from the ILL experiment results~\citep{PhysRevC.84.024617, PhysRevC.83.054615}.
We find that the background of this analysis contribution from reactor neutrinos becomes negligibly small above $10\,\mathrm{MeV}$.
In the range, it has a large spectrum shape uncertainty of  
$\sim 50\%$ from the Huber/Mueller spectrum model.
The number of reactor neutrino backgrounds is estimated to be 
$1.4\pm0.6$
including KamLAND-detector related uncertainties.
The expected number of events from the extrapolated reactor spectrum is consistent with the Daya~Bay results~\citep{An_2017} at neutrino energies of $\qty[8.125,\, 12\,\mathrm{MeV}]$, the highest energy bin in the Daya~Bay analysis.

\subsection{Accidental coincidence}
Two uncorrelated events may accidentally pass through the DC selection.
Predominantly, un-correlated long-lived spallation isotopes or radioactive decays could produce a mimic prompt event, and radioactive decays such as $^{214}$Bi and $^{208}$Tl beta/gamma-rays possibly become a 2.2\,MeV of mimic delayed event.
To estimate the random coincidence background, events were selected with appropriate prompt and delayed energies but in an off-time window of $0.2$--$1.2\,\mathrm{s}$ after the prompt event.
This off-time window is $10^3$ times larger than the antineutrino selection, providing a high statistics estimate.
The expected number of accidental coincidence background events is 
$(7.3\pm1.0) \times 10^{-2}$.

\subsection{Spallation products}
Cosmic muons induce various spallation products in KamLAND~\citep{PhysRevC.81.025807}.
Short-lived spallation products are rejected by a $2\,\mathrm{ms}$ whole volume veto. 
Some longer-lived products are a potential background in this study.
As a primary spallation cut, we apply a $2\,\mathrm{ms}$ whole volume veto for all muons in KamLAND, which has a muon rate of $\sim 0.34\,\mathrm{Hz}$.
Muons in the ID are identified when more than $\order{10^4}$ photons are detected by the PMTs.
We identify muon events in the OD when the number of OD hits exceeds $5$ or $9$ hits, before and after the OD refurbishment campaign, respectively~\citep{Ozaki:2016fmr}.

The previous analysis of KamLAND data~\citep{Gando_2012} showed that the $^{9}$Li ($\tau=257.2\,\mathrm{ms}$, $Q = 13.6\,\mathrm{MeV}$) spallation product is a challenging background.
In order to reduce this background contamination, we improved the spallation veto introducing a likelihood-ratio based muon shower tagging in addition to the primary $2\,\mathrm{ms}$ veto.

\begin{figure*}[htbp]
    \centering
    \includegraphics[width=1.0\linewidth]{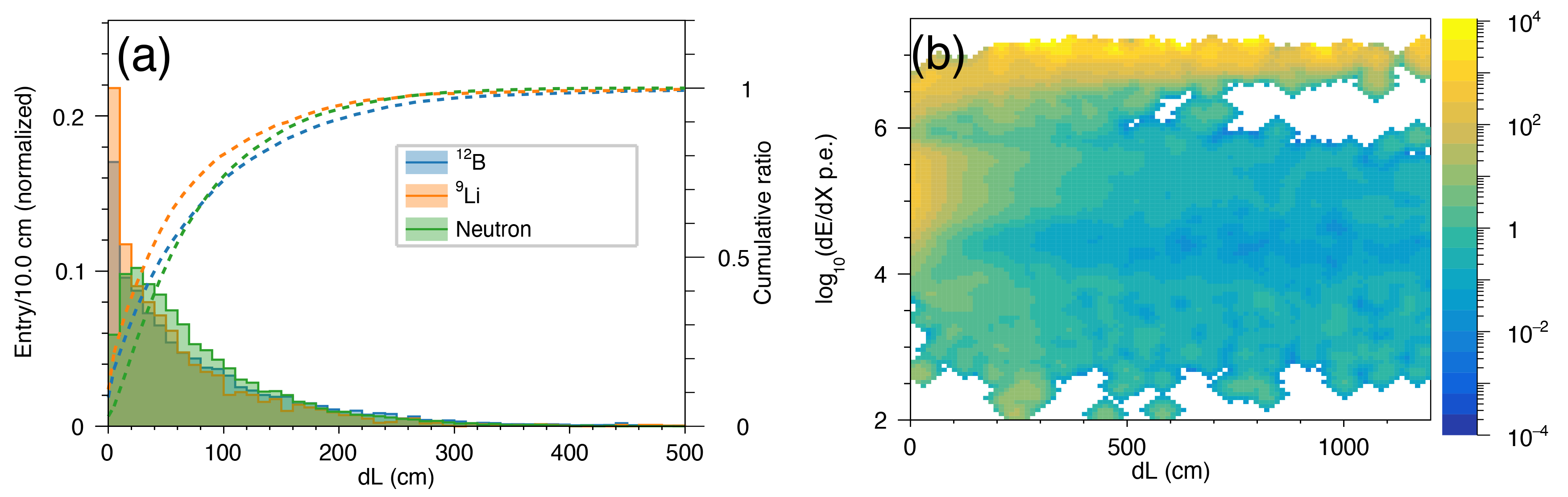}
    \caption{(a) The closest distance distribution from the muon track to the spallation products production points of $^{12}$B, $^{9}$Li, and neutron from FLUKA and \textsc{Geant4} simulations. 
    Only muons leaving more than $10^6$ photo-electrons were selected in this analysis.
    Dashed lines correspond to the cumulative ratio (right-handed scale). 
    $99\%$ of these spallation products are within $300\,\mathrm{cm}$ from the track. 
    (b) Data-driven correlation between muon charge deposition ($dE/dX$) and the closest distance of spallation product from the muon track ($dL$). The color bar shows the likelihood-ratio value.}
    \label{fig:spallationlikelihood}
\end{figure*}

Using a similar idea employed in Super-K analysis~\citep{PhysRevD.85.052007}, we evaluate a probability density function for spallation-like events taking into consideration the spatial and timing correlation of the muon track and charge deposition.
Due to low statistics of $^{9}$Li in KamLAND data with a production rate of $2.8\pm0.2\,\mathrm{kton^{-1} day^{-1}}$~\citep{PhysRevC.81.025807}, it is difficult to directly estimate the correlation between the muon track and $^{9}$Li.
Hence, instead of the $^{9}$Li, we use $^{12}$B data, whose production rate is 
$\sim 20$ times larger.
Figure~\ref{fig:spallationlikelihood}~(a) shows the closest track distance distribution ($dL$) between the muon track and the spallation production point for $^{12}$B, $^{9}$Li, and neutrons, based on a Monte-Carlo (MC) study with FLUKA~(version 2011.2x.8.patch)~\citep{BOHLEN2014211, Ferrari:2005zk} and propagated through KamLAND with \textsc{Geant4}~(version 4.9.6 patch-04)~\citep{AGOSTINELLI2003250, Geant4IEEE1610988, ALLISON2016186}.
The spread of the $^9$Li distance distribution is narrower than $^{12}$B.
This means that a tight spallation cut on the $^{12}$B data can be used to put an upper limit on the remaining $^{9}$Li background.
The difference of muon charge deposition among the spallation products was small enough.
Figure~\ref{fig:spallationlikelihood}~(b) shows the data-driven correlation between muon charge deposition per track length ($dE/dX$) and the distance of spallation products from the muon track ($dL$).
Muons depositing a large charge in the detector induce spallation products even far from the muon track.
Considering the lifetime of $^{9}$Li, a $2\,\mathrm{s}$ veto is sufficient to reject the background but the detector livetime becomes too small.
Here we define a likelihood-ratio parameter depending on the $dE/dX$, $dL$, and the time difference from the muon event to the subsequent event in order to optimize the rejection of spallation events.
To optimize the likelihood-ratio threshold while maximizing detector livetime and minimizing the spallation-cut inefficiency, we define the figure of merit (FOM) as follows~\citep{Punzi:2003bu}:
\begin{equation}
    {\mathrm{FOM} \equiv \frac{\epsilon_{\mathrm{livetime}}}{\frac{1.64}{2}+\sqrt{N_{\mathrm{non\,spall.}} \cdot \epsilon_{\mathrm{livetime}} + N_{\mathrm{spall.}} \cdot (1 - \epsilon_{\mathrm{cut}}) + N_{\mathrm{spall.}} \cdot \sqrt{(\delta_{\mathrm{cut}}^2 + \delta_{\mathrm{stat.}}^2)}}}},
\end{equation}
where $\epsilon_{\mathrm{livetime}}$ is the detector livetime ratio, $N_{\mathrm{spall.\,(non\,spall.)}}$ are the expected number of spallation (non-spallation) events without a spallation veto, $\epsilon_{\mathrm{cut}}$ is the spallation cut efficiency depending on the likelihood-ratio threshold, $\delta_{\mathrm{cut}}^2$ is the cut efficiency uncertainty, and $\delta_{\mathrm{stat.}}^2$ is the statistical uncertainty on the expected number of events.
Maximizing the FOM with the condition that the $^{9}$Li spallation cut inefficiency becomes zero consistent within the range determined from $^{12}$B, we optimize the likelihood-ratio threshold. 
This muon-shower based likelihood-ratio spallation cut allows 
$\epsilon_{\mathrm{livetime}} = 79\%$ 
of detector livetime on average and gives a spallation cut inefficiency of 
$(0.2\pm0.5)\%$, that means 99.8\% of spallation background reduction.
This gives a spallation background of $1.4\pm3.6$ in this analysis energy range.

\subsection{Fast neutrons}
The fast neutrons background comes from outside of the detector and is induced by cosmic muons in the surrounding rock and water.
Neutron scattering on protons or carbon nuclei in the liquid scintillator can mimic a prompt event.
After that, the neutrons are thermalized and captured on a proton or carbon, creating the delayed event.
The $2\,\mathrm{ms}$ veto for OD tagged muons rejects the majority of this background but some events remain due to OD inefficiency.

This background was evaluated with a \textsc{Geant4}-based MC, which included a detailed description of the KamLAND geometry (KLG4sim). 
Neutron interactions were treated with the $\texttt{QGSP\_BIC\_HP}$ physics list, while muon-nucleus interactions were activated using $\texttt{G4EmExtraPhysics}$. 
The cosmic muon directional distributions were implemented from the KamLAND spallation simulation study~\citep{PhysRevC.81.025807} which used a topological map of Mt.~Ikenoyama~\citep{IkenoyamaMap} and the MUSIC simulation tool~\citep{ANTONIOLI1997357}.
The simulated detector response in KLG4sim was tuned with various calibration data.

\begin{figure*}[htbp]
    \centering
    \includegraphics[width=1.0\linewidth]{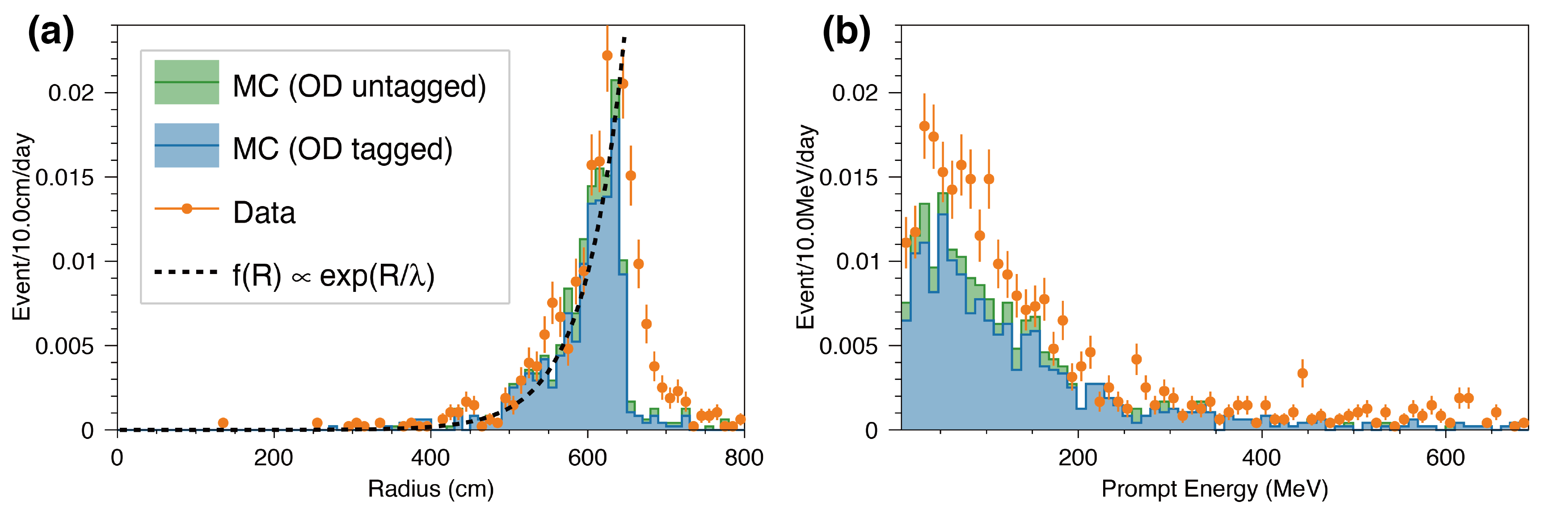}
    \caption{(a) Radial distributions of fast neutron events from MC simulation in the OD-untagged case (green) and OD-tagged case (blue) in the energy range $E_{\mathrm{prompt}} = \qty[7.5,\,30\,\mathrm{MeV}]$. 
    The black dashed line is an exponential fit to the fast neutron distribution.    
    (b) Energy spectra of fast neutron events within the $550\,\mathrm{cm}$ fiducial radius.
    The histograms are stacked.
    }
    \label{fig:fastnMC}
\end{figure*}

An equivalent of $8313$ livetime days were simulated in KLG4sim.
In the case of a muon going through the ID producing a lot of scintillation emission, the muon and associated neutrons were vetoed by the $2\,\mathrm{ms}$ whole volume veto.
Figure~\ref{fig:fastnMC}~(a) shows the reconstructed fast neutron position distribution for OD-tagged MC events.
For comparison, OD-tagged fast neutron events in the data are also shown in Figure~\ref{fig:fastnMC}~(a).
The fast neutron selection used the IBD selection described in Section~\ref{sec:antiv} except for the OD tagging and $\Delta T > 10\,\mu\mathrm{s}$ selection avoiding decay-electron contribution.
While the data has a slightly broader spread due to the difficulty of
vertex reconstruction around the boundary between liquid scintillator and buffer oil, the radial distribution in the fiducial volume are consistent between data and simulation.
The fast neutron radial distribution $f(R)$ is assumed to be $f(R) \propto \exp \qty( R/\lambda)$ as a function of the distance $R$ from the detector center, 
where $\lambda = (50.9\pm3.0)\,\mathrm{cm}$.
We use this non-uniform position distribution in the fit to data to evaluate the fast neutron background (see Section~\ref{sec:analysis}).
The fast neutron energy spectra in MC and data are shown in Figure~\ref{fig:fastnMC}~(b).

The remaining fast neutrons generate events with a few OD hits below the OD detection threshold which are therefore not tagged in the OD (OD-untagged). 
From the KLG4sim simulation, the estimated number of fast neutron background within the fiducial volume is 
$6.8 \pm 6.8$
DC pairs by scaling the detector livetime.
The uncertainty is conservatively estimated to be $100\%$, considering the poorly known production yield of neutrons in the rock. 
Events producing multiple neutrons via the ${}^{12}{\mathrm{C}}(n,\,2n){}^{11}{\mathrm{C}}$ reaction are expected to give $3.4 \pm 3.4$ DC pairs.

\subsection{Atmospheric neutrinos}
Atmospheric neutrinos produce the dominant background in this analysis via charged current (CC) and neutral current (NC) interactions.
In the CC interactions, various neutron emission modes are possible backgrounds.
Since the neutrino-nucleus interaction cross section for carbon is at least one order of magnitude smaller than for protons~\citep{KIM2009330}, the background comes mainly from proton interactions.
IBD from atmospheric electron antineutrinos is not a dominant contribution in the CC background because the mean energy is higher than our analysis energy range and small flux in this range.
Based on the atmospheric neutrino flux~\citep{PhysRevD.75.043006} and measured cross-section by MiniBooNE~\citep{PhysRevD.75.093003}, the dominant process is $\bar{\nu}_{\mu} + p \rightarrow \mu^{+} + n$.
In the KamLAND detector, the muon decay is observed as muon scintillation, muon decay, and neutron capture.
The two-prompts (muon scintillation $+$ its decay) and one-delayed (neutron capture) DC events are vetoed with $\sim78\%$ efficiency~\citep{Gando_2012}.
Its inefficiency contributes to the background.
The number of CC backgrounds is estimated to be 
$1.1 \pm 0.3$.

To estimate the NC background, we took into account the atmospheric neutrino flux~\citep{PhysRevD.75.043006}, cross sections~\citep{PhysRevD.35.785}, the neutron binding energies in carbon for the P-shell ($18.7\,\mathrm{MeV}$) and the S-shell ($41.7\,\mathrm{MeV}$) configurations and the corresponding shell populations, and de-excitation models~\citep{PhysRevD.67.076007}.
The NC interaction is given by $\nu(\bar{\nu}) + {}^{12}\mathrm{C} \rightarrow \nu(\bar{\nu}) + n + {}^{11}\mathrm{C}+\gamma$.
Most of the outgoing neutrons have a kinetic energy of less than $200\,\mathrm{MeV}$ and they scatter on protons resulting in a visible energy of typically less than $100\,\mathrm{MeV}$.
The details of the background signatures and estimations are described in \citet{Gando_2012}.
The resulting expected number of NC interactions in this data set is 
$20.6 \pm 5.9$,
where the uncertainty comes from the atmospheric neutrino flux and the cross section 
which are combined to provide $\sim30\%$ in total.

For comparison, we also estimate the atmospheric neutrino background with NEUT~(version 5.4.0.1)~\citep{Hayato:2009zz, Hayato:2021heg}.
The interaction models are summarized in Table~\ref{tab:neutsim}.
We use the \citet{PhysRevD.92.023004} model for the atmospheric neutrino flux including the matter oscillation effect implemented in the Prob3++~\citep{Prob3}.
The de-excitation model for oxygen is incorporated in NEUT, but that for carbon is missing.
After the final state interaction in NEUT, the outgoing particles were introduced into KLG4sim.
The response of this simulation was compared to KamLAND data in the $200\,\mathrm{MeV}$--$1.5\,\mathrm{GeV}$ energy range, outside of the fast neutron background.
Although there are large uncertainties from the fast neutrons in the $30$--$200\,\mathrm{MeV}$ energy range, data and MC are consistent within the errors.
Below $100\,\mathrm{MeV}$, the NC quasi-elastic scattering (NCQE) is dominant. 
The NC two-particle-two-hole (NC~2p2h) interaction will also contribute, but assuming that the ratio of the NC~2p2h to the NCQE cross sections is similar to the corresponding the CC ratio, roughly $5$--$10\%$~\citep{PhysRevC.83.045501}, the contribution is estimated to be small compared to the NCQE contribution and its large uncertainties.

For the DC energy selection of $7.5$--$30\,\mathrm{MeV}$ in NEUT based estimation, the remaining CC and NC backgrounds are estimated to be 
$0.9^{+0.3}_{-0.4}$ and $16.5^{+5.1}_{-4.5}$,
respectively.
The NEUT background estimate are smaller, but are consistent within the uncertainties.
The neutron multiplicity in the NC reaction may play a role in the models and affect the background estimate due to the DC requirement of selecting a single neutron capture.
The energy spectrum shape is concrete on the de-excitation models of carbon and the proton-scattering by neutrons in the numerical calculation, but this simulation-based estimation does not include the de-excitation.
Therefore we took the numerically calculated spectrum~\citep{Gando_2012} and estimated the number of NC backgrounds to be $20.6\pm5.9$ in this analysis.
We treat the number of NC background events as a free parameter in this analysis, independent the number of backgrounds from the estimation models, and use the energy spectrum to constrain.

\begin{deluxetable*}{ll}[htbp]
    \label{tab:neutsim}
    \tablecaption{The atmospheric neutrino background study used the following nuclear interaction models in NEUT.}
    \tablewidth{0pt}
    \tablehead{
        \colhead{Interaction} & \colhead{Reference model}
    }
    \startdata
        NCQE nuclear model & \begin{tabular}{l} \citet{Ankowski:2011ei}  \end{tabular}\\
        CCQE nuclear model & \begin{tabular}{l} \citet{Gran:2013kda}  \end{tabular}\\
        Axial vector mass for quasi elastic & \begin{tabular}{l}$M_{\mathrm{A}}^{\mathrm{QE}} = 1.2\,\mathrm{GeV\,c^{-2}}$  \end{tabular}\\
        Fermi momentum (NCQE) & \begin{tabular}{l} $217\,\mathrm{MeV\,c^{-1}}$  \end{tabular}\\
        Two-nucleon scattering (2p2h) \begin{tabular}{l} (NC)\\ (CC)\end{tabular} & \begin{tabular}{l} Not treated\\ Nieves~\citep{PhysRevC.83.045501} \end{tabular} \\
        Vector form factor (NCQE/CCQE) & \begin{tabular}{l}BBBA05~\citep{BRADFORD2006127}  \end{tabular}\\
        Axial vector form factor & \begin{tabular}{l}\citet{Graczyk:2007bc} and \citet{Nowak:2009se}  \end{tabular}\\
        Single pion production & \begin{tabular}{l}\citet{Berger:2007rq}  \end{tabular}\\
        Deep inelastic scattering & \begin{tabular}{l}GRV98 parton distribution~\citep{Gluck:1998xa}\\ \,\, with Bodek-Yang corrections~\citep{doi:10.1063/1.1594324}\end{tabular} \\
        Final state interaction   & \begin{tabular}{l} \citet{Hayato:2021heg}  \end{tabular}\\
    \enddata
\end{deluxetable*}

\section{Analysis and Results} \label{sec:analysis}
Our search for astrophysical electron antineutrino signals
fitted the energy spectra and radial distributions in data to the estimated backgrounds. 
The fast neutron background contributes with a large uncertainty but is mostly concentrated at the outer radius, while the other backgrounds and neutrino candidates have a uniform distribution in the detector.
The atmospheric neutrino NC interaction is the primary background in this analysis.
We used the following $\chi^2$ to fit the number of atmospheric NC backgrounds and the number of astrophysical neutrinos:
\begin{eqnarray}
    \label{eq:chi2def}
    \chi^2 &=& \chi^2_{\mathrm{rate}} + \chi^2_{\mathrm{shape}} + \chi^2_{\mathrm{penalty}} + \chi^2_{\mathrm{BG}}, 
\end{eqnarray}
with,
\begin{equation}
    \chi^2_{\mathrm{rate}} = \frac{\qty(N_{\mathrm{observed}} - N_{\mathrm{astro.\,\nu}} - N_{\mathrm{NC}} - \sum_i^5 N_{\mathrm{BG_i}})^2}{\sigma_{\mathrm{stat.}}^2},
\end{equation}
\begin{equation}
    \chi^2_{\mathrm{shape}} = \sum^{N_{\mathrm{observed}}}_{n} \qty{ -2\ln\qty( \frac{\sum_j^7 N_{j} f_j(R) \cdot g_j(E)}{ \sum_j^7 N_j} ) }, 
\end{equation}
\begin{equation}
    \chi^2_{\mathrm{penalty}} = \sum_k \delta^2_{\mathrm{k}}, 
\end{equation}
\begin{equation}
    \chi^2_{\mathrm{BG}} = \sum_i^5 \frac{(N_{\mathrm{BG_i}}-N_{\mathrm{BG_i}}^{\mathrm{expected}})^2}{\delta^2_{\mathrm{BG_i}}}, 
\end{equation}
where $N_{\mathrm{observed}}$ is the number of observed IBD candidates, $N_{\mathrm{astro.\nu}}$ is the number of astrophysical neutrino events, $N_{\mathrm{NC}}$ is the number of atmospheric neutrino NC background events, $N_{\mathrm{BG_i}} (i=1,2,...5)$ represent the number of the other background contributions (see Table~\ref{tab:fitresult}).
The statistical uncertainty $\sigma_{\mathrm{stat.}}$ is the square root of the total number of expected events. 
In the shape $\chi^2$ term ($\chi^2_{\mathrm{shape}}$), $R$ is the radius, $E$ is the energy, $f_j(R)$ is the normalized radius distribution, and $g_j(E)$ the normalized energy spectrum for each contribution $j$ where $j=1,2,...7$ correspond to the astrophysical neutrino signal, atmospheric NC background, and the other 5 background contributions. 
We use them as an unbinned log-likelihood fit to test $\chi^2_{\mathrm{shape}}$.
Only fast neutrons have an exponential radius distribution $f_i(R)$, the other contribution are uniform.
We integrate the energy and radius over
$7.5$--$30\,\mathrm{MeV}$ and $0$--$550\,\mathrm{cm}$, respectively.
The penalty term ($\chi^2_{\mathrm{penalty}}$) is computed from the systematic uncertainties $(\delta_k)$: energy spectrum shape uncertainty, radial distribution uncertainty, detector efficiency uncertainty, and energy scale uncertainty.
In the background term ($\chi^2_{\mathrm{BG}}$), $N_{\mathrm{BG_i}}$ is the number of the $i$-th background events, $N_{\mathrm{BG_i}}^{\mathrm{expected}}$ is the expected number of the $i$-th background component, and $\delta^2_{\mathrm{BG_i}}$ is its associated uncertainty.

\subsection{Solar electron antineutrino}
\begin{figure}[htbp]
    \centering
    \includegraphics[width=0.8\linewidth]{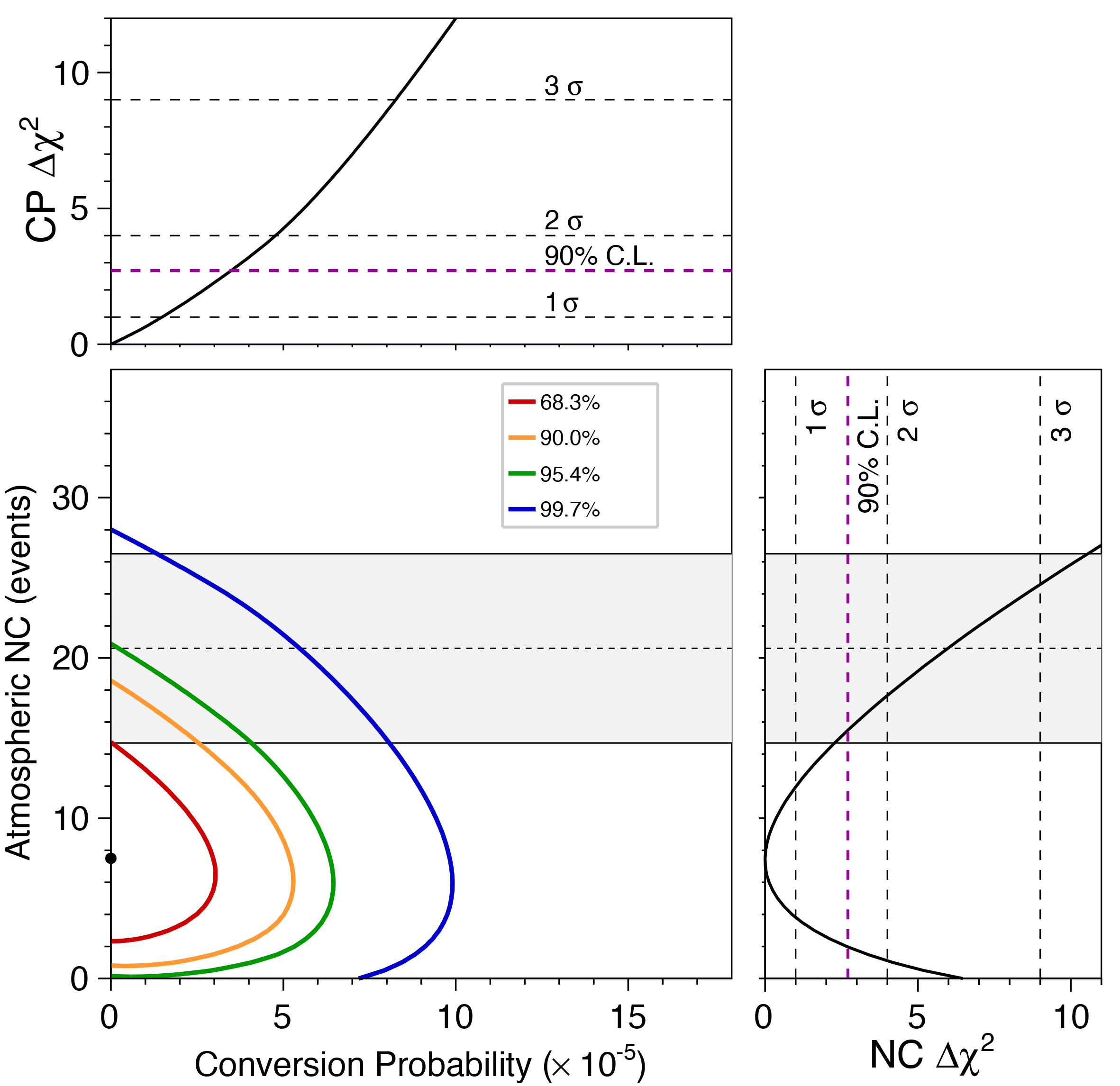} 
    \caption{The results and allowed regions for the solar $\nu_e \rightarrow \bar{\nu}_e$ conversion probability and the number of atmospheric neutrino NC interactions. 
    Color contours correspond to $1\sigma$ (red), $90\%$ (orange), $2\sigma$ (green), and $3\sigma$ (blue).  
    The best-fit conversion probability and NC events are $0$ and $7.5$, respectively (black dot). 
    The horizontal hatched region represent the expected number of NC events with $1\sigma$ uncertainty. 
    Top and right panels are 1-dimensional $\Delta \chi^2$ distributions for conversion probability (CP) and number of atmospheric neutrino NC interaction, respectively.
    The upper limit on the conversion probability is 
    $3.5 \times 10^{-5}$ at $90\%$~CL.
    }
    \label{fig:resultsolar}
\end{figure}

\begin{figure*}[htbp]
    \centering
    \includegraphics[width=1.0\linewidth]{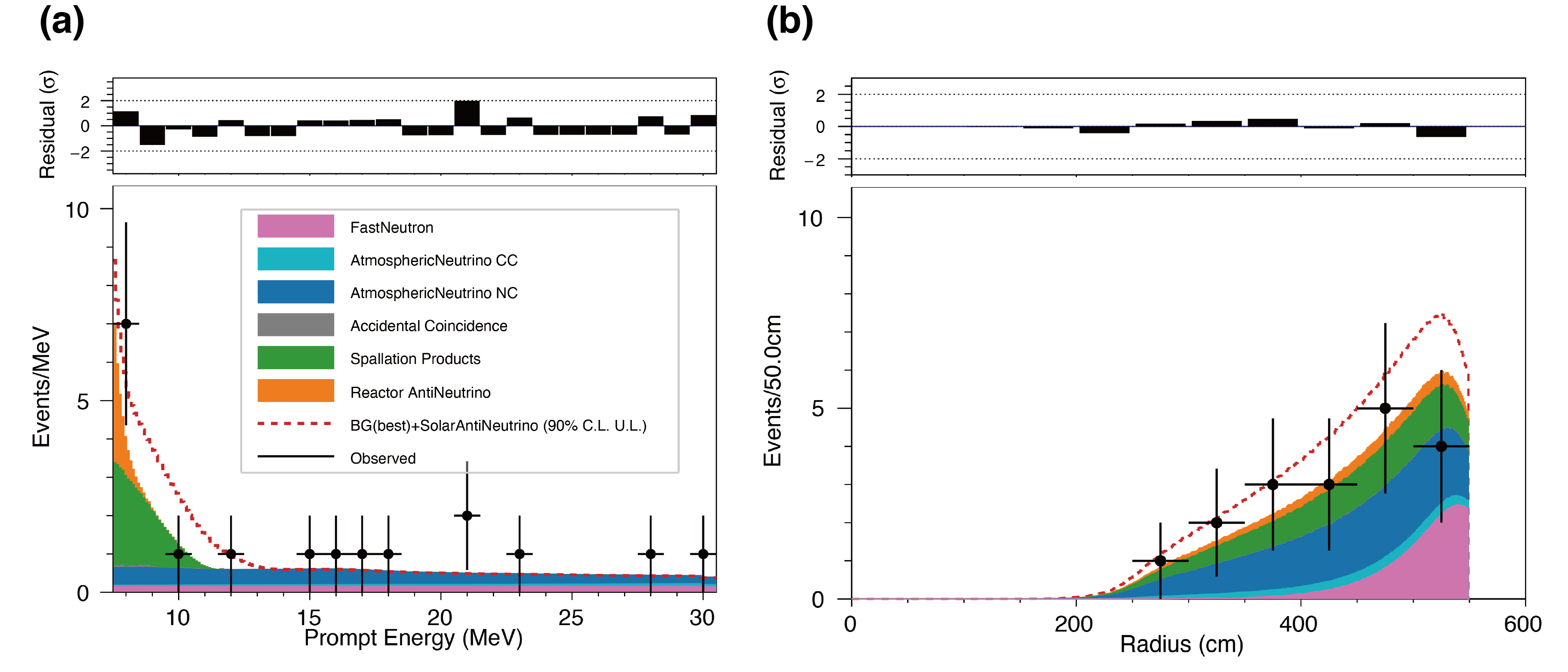}
    \caption{Results for the solar $\bar{\nu}_e$ fit for (a) the prompt energy spectrum and (b) the radial distribution. 
    The filled histograms are the best-fit background contributions.
    The red dashed lines show the $90\%$~CL upper limit for solar $^{8}$B $\bar{\nu}_e$. 
    All histograms are stacked. 
    The upper panels show residuals.
    }
    \label{fig:fitspectrumsolar}
\end{figure*}

Assuming an unoscillated $^{8}$B neutrino flux of $5.94 \times 10^{6}\,\mathrm{cm^{-2}s^{-1}}$~\citep{PenaGaray:2008qe}, the region allowed by the fit is shown in Figure~\ref{fig:resultsolar} and summarized in Table~\ref{tab:fitresult}.
The best fit values for the $\nu_e \rightarrow \bar{\nu}_e$ conversion probability and NC events are 
$0.0$ and $7.5 \pm 3.4$, respectively.
The number of atmospheric neutrino NC interactions is smaller than the estimate but model $2\sigma$ and data $2\sigma$ bands overlap.
This value is also consistent with the NEUT simulation result within $1\sigma$. 
Figure~\ref{fig:fitspectrumsolar} shows the energy and radial distributions for best-fit backgrounds and the upper limit for solar $^{8}$B $\bar{\nu}_e$ with $90\%$ confidence level (CL).
All residual values are within $\pm2\sigma$ region. 
The obtained upper limit on the conversion probability is  
$3.5 \times 10^{-5}$ at $90\%$~CL, corresponding to a $60\,\mathrm{cm^{-2}s^{-1}}$ 
solar ${}^{8}$B $\bar{\nu}_e$ flux limit above $8.3\,\mathrm{MeV}$ of neutrino energy 
(containing $30\%$ of the solar $^{8}$B neutrino flux).
In a comparable case of using a measurement $^{8}$B neutrino flux of $5.25\times 10^{6}\,\mathrm{cm^{-2}s^{-1}}$~\citep{PhysRevC.88.025501}, the upper limit on the conversion probability becomes $3.9 \times 10^{-5}$ at $90\%$~CL.
This result improves on the previous KamLAND study~\citep{Gando_2012} and is the most stringent upper limit to date.

From the upper limit on the conversion probability and Equation~(\ref{eq:rsftprob}), we also obtain the upper limit on the neutrino magnetic moment ($\mu$) and the transverse solar magnetic field ($B_{T}$) in the region of neutrino production:
\begin{equation}
    \mu < 4.9\times10^{-10} \mu_B \qty(\frac{10\,\mathrm{kG}}{B_{T}(0.05R_{\odot})}),
\end{equation}
using $34^{\circ}$ for the mixing angle $\theta_{12}$~\citep{PhysRevD.83.052002}.
This bound is weaker than the most stringent upper limit of $0.28 \times 10^{-10} \mu_B$ from the solar neutrino spectrum measurement by Borexino~\citep{PhysRevD.96.091103}.

\begin{deluxetable*}{lllh}[htbp]
    \label{tab:fitresult}
    \tablecaption{Summary of estimated backgrounds and best fit parameters.}
    \tablewidth{0pt}
    \tablehead{
        \colhead{} & \colhead{Expected} & \colhead{Best fit} & \nocolhead{Fit condition} 
    }
    \startdata
    Reactor              & $1.4  \pm 0.6$   & \begin{tabular}{l} $1.3$ \end{tabular} & limited \\
    Accidental           & $(7.3 \pm 1.0)\times 10^{-2}$ & \begin{tabular}{l} $7.3 \times 10^{-2}$ \end{tabular} & limited \\
    Fast neutron         & $6.8  \pm 6.8$   & \begin{tabular}{l} $3.3$ \end{tabular} & limited \\
    Spallation           & $1.4  \pm 3.6$   & \begin{tabular}{l} $4.5$ \end{tabular} & limited \\
    Atmospheric-$\nu$ NC &  $20.6  \pm 5.9$ & \begin{tabular}{l} $7.5$ \end{tabular} & free (scan) \\
    Atmospheric-$\nu$ CC & $1.1  \pm 0.3$   & \begin{tabular}{l} $1.1$ \end{tabular} & fixed \\
    \hline
    Solar $^{8}$B $\bar{\nu}_e$ & N/A         & \begin{tabular}{l} $0$ (best)\\ $5.9$ ($90\%$~CL upper limit)\end{tabular} & free (scan) \\ \hline
    Total                & $31.4 \pm 9.7$   & \begin{tabular}{l} $17.8$ (best)\\ $23.7$ ($90\%$~CL upper limit)\end{tabular} & {} \\ \hline
    Observed             & \multicolumn{2}{l}{$18$} & {} \\
    \enddata
\end{deluxetable*}

\subsection{Supernova relic neutrinos}
To search for SRNs, we fit with different theoretical models that produce $\bar{\nu}_e$ emission: the Kaplinghat model~\citep{PhysRevD.62.043001} (Kaplinghat+00), the Horiuchi model in the case of $6\,\mathrm{MeV}$ effective temperature~\citep{PhysRevD.79.083013} (Horiuchi+09, $6\,\mathrm{MeV}$), the Nakazato maximum model in the case of inverted mass ordering (Nakazato+15 max, IH), and the Nakazato minimum model in the case of normal mass ordering (Nakazato+15 min, NH)~\citep{Nakazato_2013, Nakazato_2015}.
From the $\chi^2$ defined in Equation~(\ref{eq:chi2def}), we find no significant excess of SRNs with any of the models.
As an example of the fitting result with the Nakazato+15~(max, IH) model, the best-fit value for the number of SRNs is $0$ events while the number of NC backgrounds is $7.5$ events.
This result is consistent with the calculated number of SRN events of 
$0.4$
in KamLAND. 
The $90\%$~CL upper limit on the number of events is 
$9.3$.
The upper flux limit is calculated to be $108\,\mathrm{cm^{-2}s^{-1}}$ from
\begin{equation}
    F_{90} = N_{90} \times \frac{\int_{E_{\mathrm{min}}}^{E_\mathrm{max}} \qty(\dv{F}{E})_M dE}{\int_{E_{\mathrm{min}}}^{E_\mathrm{max}} \qty(\dv{N}{E})_M dE},
\end{equation}
where $F_{90}$ and $N_{90}$ are the upper limits on flux and the number of events, respectively.
The $\qty(\dv{F}{E})_M$ and $\qty(\dv{N}{E})_M$ are the theoretical differential flux and spectrum, respectively, for the SRN models.
This $90\%$~CL flux upper limit is still much higher than the expected flux of 
$5.1\,{\mathrm{cm^{-2}s^{-1}}}$.
Table~\ref{tab:fitresultsrn} shows a summary of the fit results for each theoretical model and corresponding upper limit.
For all tested models, the best fit number of SRN is $0$, and NC background is $7.5$. 
However, the reported upper limit changes for each model due to differences in the underlying theoretical energy spectrum.

\begin{deluxetable*}{ccccc}[htbp]
    \label{tab:fitresultsrn}
    \tablecaption{Summary of obtained SRN flux and number of event upper limits ($90\%$~CL).}
    \tablewidth{0pt}
    \tablehead{
        \colhead{Model} & \colhead{} & \colhead{$N_{90}$ (event)} & \colhead{$F_{90}$ ($\mathrm{cm^{-2}s^{-1}}$)} & \colhead{Expected flux ($\mathrm{cm^{-2}s^{-1}}$)} 
    }
    \startdata
    Kaplinghat+00 & \citep{PhysRevD.62.043001} & $9.4$ & $74.5$ & $19.9$ \\ 
    Horiuchi+09 ($6\,\mathrm{MeV}$) & \citep{PhysRevD.79.083013} & $10.2$ & $61.6$ & $5.8$ \\  
    Nakazato+15 (max, IH) & \citep{Nakazato_2013, Nakazato_2015} & $9.3$ & $108$ & $5.1$ \\ 
    Nakazato+15 (min, NH) & \citep{Nakazato_2013, Nakazato_2015} & $8.9$ & $105$ & $2.2$    
    \enddata
\tablecomments{$F_{90}$ and $N_{90}$ are the $90\%$~CL upper limits of flux and number of events, respectively. The expected flux is integrated over our analysis energy range $E_{\mathrm{prompt}}=\qty[7.5,\,30\,\mathrm{MeV}]$.}
\end{deluxetable*}

\subsection{Model independent flux}
\begin{figure}[htbp]
    \centering
    \includegraphics[width=1.0\linewidth]{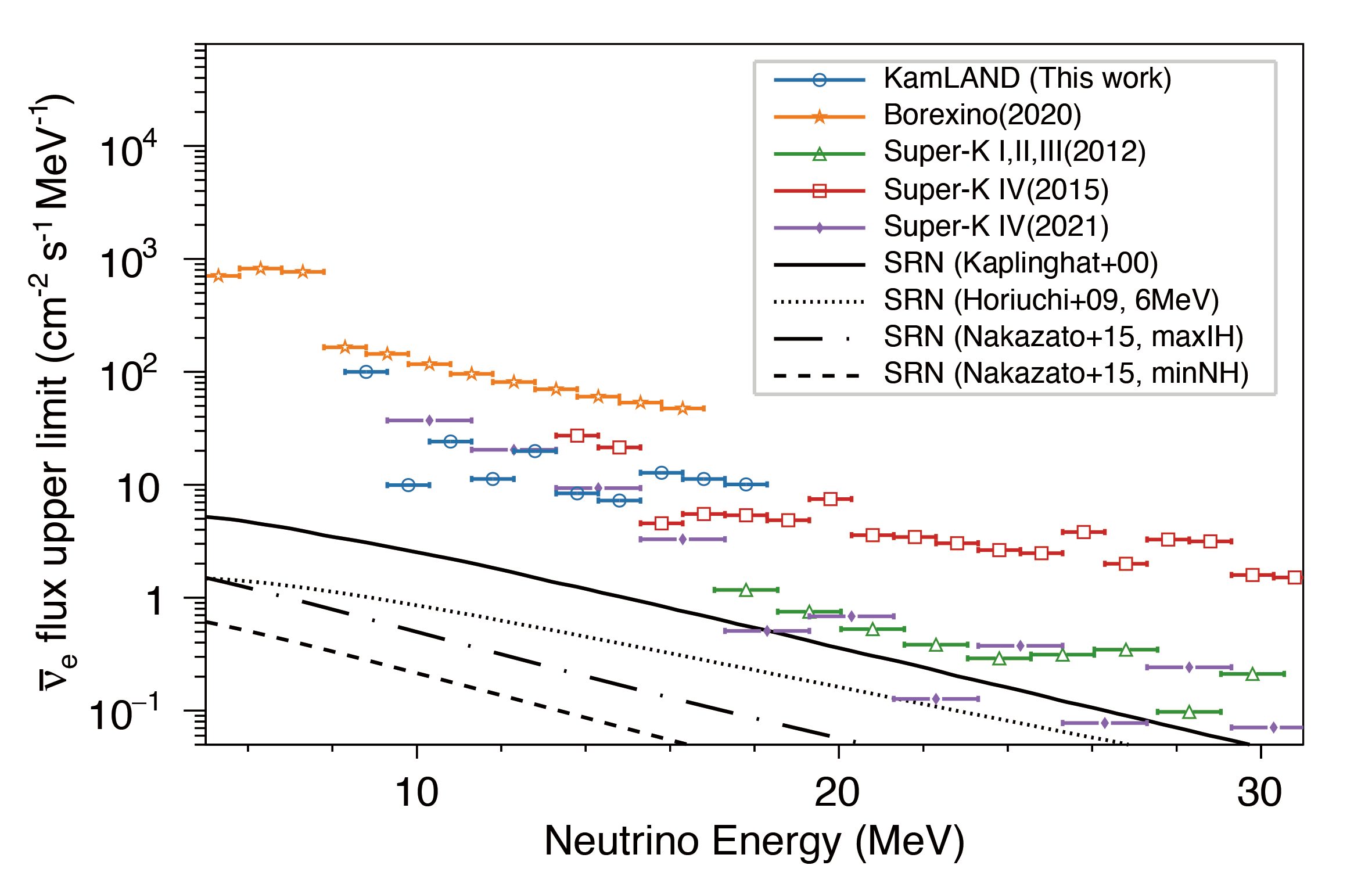} 
    \caption{
    Model-independent upper limits on the $\bar{\nu}_e$ flux (at $90\%$~CL). 
    This work is compared to Borexino~\citep{AGOSTINI2021102509} in the case including atmospheric neutrino background, Super-K I/II/III~\citep{PhysRevD.85.052007}, Super-K IV~\citep{ZHANG201541}, and Super-K IV~\citep{kamiokandecollaboration2021diffuse}. 
    The black lines show different theoretical SRN fluxes.
    }
    \label{fig:modelindependentflux}
\end{figure}

We also present model-independent upper limits on the $\bar{\nu}_e$ flux assuming monochromatic neutrino energies.
The flux upper limits ($\phi_{90}$) are calculated with
\begin{equation}
    \phi_{90} = \frac{N_{90}}{N_p \cdot \sigma \cdot \epsilon_{\mathrm{IBD}} \cdot T},
\end{equation}
where $N_{90}$ is the $90\%$~CL upper limit on the number of $\bar{\nu}_e$ in a $1\,\mathrm{MeV}$ wide bin using the \citet{PhysRevD.57.3873} approach, $N_p$ is the number of target protons, $\sigma$ is the IBD reaction cross section, $\epsilon_{\mathrm{IBD}}$ is the detector efficiency, and $T$ is the detector livetime.
Figure~\ref{fig:modelindependentflux} shows the resulting electron antineutrino flux in comparison with results from Borexino~\citep{AGOSTINI2021102509}, Super-K~\citep{PhysRevD.85.052007,ZHANG201541,kamiokandecollaboration2021diffuse}, and various theoretical SRN models~\citep{PhysRevD.62.043001,PhysRevD.79.083013,Nakazato_2013,Nakazato_2015}.
While our results do not yet exclude SRN models, they provide the strictest flux limits for $E_{\nu}=\qty[8.3,\,13.3\,\mathrm{MeV}]$.
Table~\ref{tab:modelindependentflux} shows a summary of the flux upper limits per a bin.

\begin{deluxetable*}{cc}[htbp]
    \label{tab:modelindependentflux}
    \tablecaption{The obtained $\bar{\nu}_e$ upper flux limit ($90\%$~CL) assuming all $\bar{\nu}_e$ have a monochromatic energy.}
    \tablewidth{0pt}
    \tabletypesize{\scriptsize}
    \tablehead{
        \colhead{Energy ($\mathrm{MeV}$)} & \colhead{Flux upper limit at $90\%$~CL ($\mathrm{cm^{-2} s^{-1} MeV^{-1}}$)} 
    }
    \startdata
    $8.3$--$9.3$   & $98.1$ \\
    $9.3$--$10.3$  & $9.5$ \\
    $10.3$--$11.3$ & $23.8$ \\
    $11.3$--$12.3$ & $11.2$ \\
    $12.3$--$13.3$ & $19.8$ \\
    $13.3$--$14.3$ & $8.4$ \\
    $14.3$--$15.3$ & $7.3$ \\
    $15.3$--$16.3$ & $12.8$ \\
    $16.3$--$17.3$ & $11.2$ \\
    $17.3$--$18.3$ & $10.1$
    \enddata
\end{deluxetable*}

\subsection{Dark matter self annihilation}
\begin{figure}[htbp]
    \centering
    \includegraphics[width=0.8\linewidth]{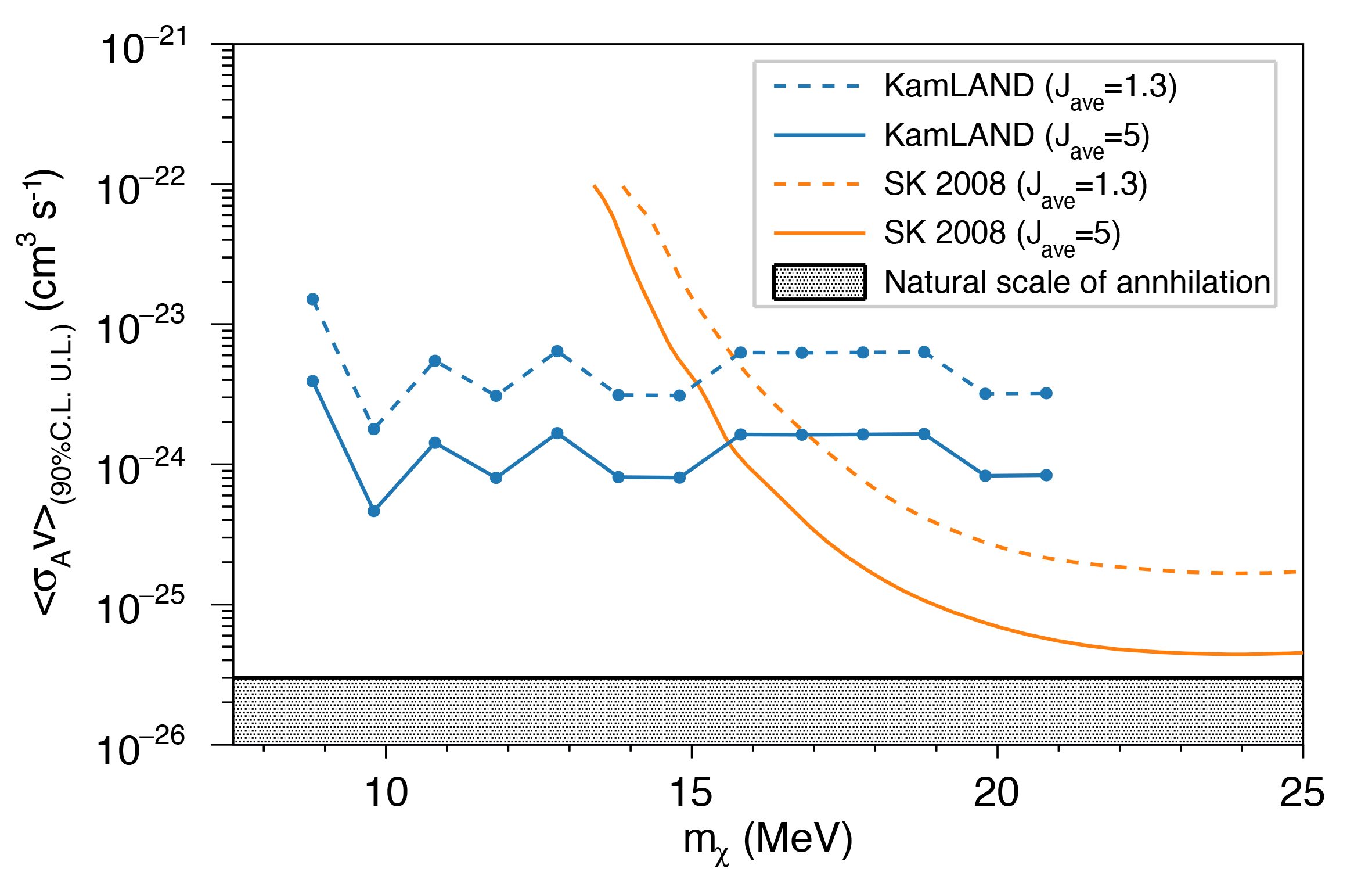} 
    \caption{Upper limits on the dark matter self-annihilation cross section at $90\%$~CL from KamLAND and Super-K~\citep{PhysRevD.77.025025}. 
    Two benchmark cases for the angular-averaged intensity $\mathcal{J}_{\mathrm{ave}}$ are shown, $\mathcal{J}_{\mathrm{ave}} = 1.3$ (dashed line) and $5.0$ (solid line).
    The shadowed region corresponds to the natural scale of the annihilation cross section as $3\times10^{-26}\,\mathrm{cm}^3\,\mathrm{s}^{-1}$~\citep{PhysRevD.86.023506}.}
    \label{fig:resultdarkmatter}
\end{figure}

The $\bar{\nu}_e$ flux upper limit per energy bin can be translated to a dark matter self-annihilation cross section  limit~\citep{PhysRevD.77.025025}.
From Equation~(\ref{eq:dmcrosssection}), we obtain an upper limit of $\langle \sigma_A \vb{v} \rangle < (1$--$11) \times 10^{-24}\,{\mathrm{cm^{-3}s^{-1}}}$ ($90\%$~CL) for the benchmark case of $\mathcal{J}_{\mathrm{ave}}=1.3$ (see Figure~\ref{fig:resultdarkmatter}).
This result is the most stringent constraint on the self-annihilation cross section for $m_{\chi} < 15\,\mathrm{MeV}$.

\section{Summary} \label{sec:summary}
We searched for astrophysical $\bar{\nu}_e$ in the neutrino energy range $8.3$ to $30.8\,\mathrm{MeV}$ with $4528.5$ livetime days of KamLAND data.
No significant excess was found over the expected backgrounds.
We presented the strictest upper limit on the conversion probability of solar $^{8}$B neutrinos to antineutrinos, 
$3.5\times10^{-5}$ (at $90\%$~CL).
Assuming various model predictions, the upper limit on the SRN flux translates to $60$--$110\,\mathrm{cm^{-2} s^{-1}}$.
We also give the strictest upper limit on the model independent flux below $13.3\,\mathrm{MeV}$ but this limit is still an order of magnitude larger than SRN model predictions.
The upper limits on the dark matter self-annihilation cross-section to neutrino pairs are the most stringent for dark matter particle masses below $15\,\mathrm{MeV}$.
Our results for the model-independent flux limit (Table~\ref{tab:modelindependentflux}) can set limits on various astrophysical $\bar{\nu}_e$'s, for instance, neutrinos from sterile neutrino decay~\citep{Hostert:2020oui} and primordial black hole dynamics~\citep{PhysRevLett.125.101101, PhysRevD.103.043010,calabrese2021primordial}.

Further background suppression is necessary to improve the solar $^{8}$B $\bar{\nu}_e$ and SRN sensitivity.
A future neutrino detector at a deep underground site will suppress the spallation background~\citep{PhysRevD.99.012012, Guo_2021}.
A larger distance to nuclear power plants will reduce the reactor neutrino component.
More detailed measurements of the high-energy reactor neutrino spectrum are necessary, including the end point.
The background arising from atmospheric neutrinos is the most challenging.
Pulse shape discrimination (PSD) may reduce this contribution~\citep{LI2016303} in a future large neutrino detector such as JUNO~\citep{An_2016}.
Although the fast scintillation decay time of the current KamLAND liquid scintillator cocktail and significant re-emission, PSD can be improved by the detector upgrades and requires excellent timing resolution for PMTs.
The KamLAND2 detector upgrade program intends to use a linear-alkyl-benzene based liquid scintillator which would realize the PSD due to a slower scintillation decay time compared to the current KamLAND liquid scintillator~\citep{Asakura:2014lma, 10.1093/ptep/ptz064, Yuto:2020euy, Nakamura_2020, Takeuchi:2020cme}.

\begin{acknowledgments}
The KamLAND experiment is supported by 
JSPS KAKENHI Grants 
18J10498, 
19H05102, 
and 19H05803; 
the World Premier International Research Center Initiative (WPI Initiative), MEXT, Japan; 
Netherlands Organization for Scientific Research (NWO); 
and under the U.S. Department of Energy (DOE) Contract 
No.~DE-AC02-05CH11231,
the National Science Foundation (NSF) No.~NSF-1806440, 
NSF-2012964, 
as well as other DOE and NSF grants to individual institutions.  
The Kamioka Mining and Smelting Company has provided service for activities in the mine.  
We acknowledge the support of NII for SINET4. 
We also thank Y.~Hayato for advising our atmospheric neutrino simulation with NEUT.
This work is partly supported by 
the Graduate Program on Physics for the Universe (GP-PU). 
\end{acknowledgments}

\bibliographystyle{aasjournal}
\bibliography{main}{}

\end{document}